\documentclass[aps,prl,amsmath,amssymb,twocolumn,superscriptaddress]{revtex4}
\pdfoutput=1

\usepackage{graphicx}% Include figure files
\renewcommand{\d}{\mathrm{d}}
\newcommand{\ii}{\mathrm{i}}

\newcommand{\mr}[1]{\mathrm{#1}}
\newcommand{\refcite}[1]{Ref.~\onlinecite{#1}}

\begin{document}

\title{Fully Overheated Single-Electron Transistor}

\author{M.~A.~Laakso}
\email[]{matti.laakso@tkk.fi}
\author{T.~T. Heikkil\"a}
\affiliation{Low Temperature Laboratory, Helsinki University of Technology, P.O.~Box 5100 FIN-02015 TKK, Finland}
\author{Yuli V.~Nazarov}
\affiliation{Kavli Institute of Nanoscience, Delft University of Technology, 2628 CJ Delft, The Netherlands}
\date{\today}

\begin{abstract}
We consider the fully overheated single-electron transistor, where the heat balance is determined entirely by electron transfers. We find three distinct transport regimes corresponding to cotunneling, single-electron tunneling, and a competition between the two. We find an anomalous sensitivity to temperature fluctuations at the crossover between the two latter regimes that manifests in an exceptionally large Fano factor of current noise.
\end{abstract}


\maketitle

\begin{figure}
 \includegraphics[width=\columnwidth]{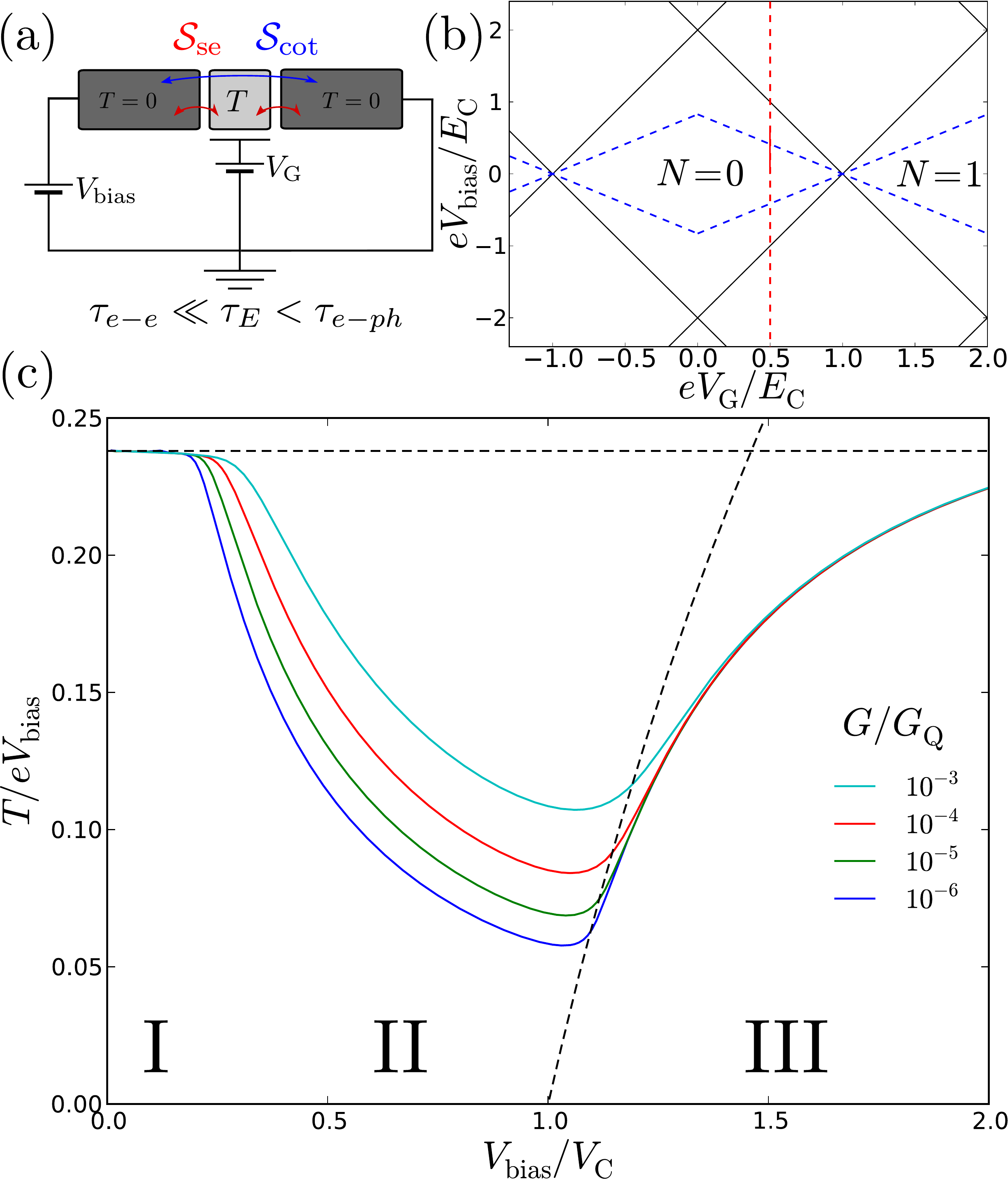}
 \caption{(color online) (a) SET biased by a voltage $V_\mr{bias}$. The charge in the central island can be tuned with the gate voltage $V_\mr{G}$. (b) Coulomb diamonds in a symmetric SET. Blue dashed lines show the threshold voltage $V_\mr{C}$ for SE tunneling. (c) Average temperature of the island versus bias voltage along the red vertical line in (b) for various tunnel conductances $G$. This illustrates the three transport regimes: I cotunneling, II competition, III SE tunneling. Dashed lines are asymptotes to pure cotunneling and pure SE tunneling.}
 \label{fig:schema}
\end{figure}

A single-electron transistor (SET) \cite{averin86}, shown schematically in Fig.~\ref{fig:schema}(a), is one of the most thoroughly studied and widely used nanodevices. It has found its way to numerous applications in thermometry \cite{pekola94}, single-electron pumping \cite{pothier92}, charge detection \cite{schoelkopf98,devoret00}, and detection of nanoelectromechanical motion \cite{knobel03, lahaye04}. The current noise in a SET limits the measurement sensitivity and is thus worth investigating \cite{korotkov94b}.

Nanodevices of sufficiently small size are overheated: the electron temperature in the device deviates from the lattice temperature. The temperature may fluctuate in this regime \cite{heikkila09}, thereby affecting the current noise in the device. This motivates us to study overheating under Coulomb blockade conditions. In this Letter we concentrate on a fully overheated SET. We assume the electron--phonon relaxation time, $\tau_{e-ph}$, to exceed by far the electron dwell time, so that the temperature is determined entirely from the balance of electronic heat transfers. In addition, we restrict our study to a symmetrically biased SET with junction conductance $G \ll G_\mr{Q}\equiv e^2/\pi\hbar$ and a vanishing temperature of the leads.

An early paper \cite{korotkov94} addressed overheating in a SET in the regime of single-electron (SE) tunneling. It has been found that overheating instigates the SE transport at the threshold voltage $V_\mr{C} = (\sqrt{2}-1)V_\mr{th}$, that is, well below the zero-temperature Coulomb blockade threshold $V_\mr{th}=2\delta E^+/e$, ($\delta E^\pm=E_\mr{C}\mp eV_\mr{G}$ is the charging energy, $E_\mr{C}=e^2/2C$) as shown in Fig.~\ref{fig:schema}(b). We complement the consideration with electron cotunneling \cite{averin90}, which modifies the picture rather radically. We recognize that single-electron processes below $V_\mr{C}$ try to cool the island, competing with the electron--hole excitations left behind by inelastic cotunneling that heat it up. This gives rise to a new transport regime: {\it competition} regime. The three regimes are evident in the voltage dependence of temperature as shown in Fig.~\ref{fig:schema}(c). At low voltage cotunneling dominates and the temperature scales with voltage, $k_\mr{B}T/eV_\mr{b}\approx0.24$, as expected for a fully overheated nanodevice. Sufficiently high temperature activates SE transfers that cool the island and sets the competition in. The temperature/voltage ratio reaches the minimum $k_\mr{B}T_\mr{C}/eV_\mr{C} = 1/\sqrt{2}\ln(\mr{const.}\times G_\mr{Q}/G)$ near $V_\mr{C}$. Above the threshold, the SE processes heat the island resulting in an increase of temperature. A pure SE picture captures only this rise predicting $T=0$ at $V_\mr{b}<V_\mr{C}$.

The most interesting features can be found at $V_\mr{b}\approx V_\mr{C}$ where the crossover between competition and SE regimes takes place. We show that near the crossover the electric current is anomalously sensitive to temperature changes: it is significantly modified by a temperature change $\delta T \ll T$. The underlying mechanism of the anomalous sensitivity is the strong temperature dependence of thermally activated tunneling rates. The overheated SET also detects fluctuations of its own temperature, manifest in an enhanced current noise. The current noise $S_I$ is commonly characterized by Fano factor $F \equiv S_I/2eI$, $F\leq1$ for most of nanodevices. At the crossover, $F \propto \ln^4(G_\mr{Q}/G)$ reaching an impressive $1500$ for $G/G_\mr{Q}=10^{-6}$.

We implement a method that allows to access the full statistics of temperature and current fluctuations. The statistics are described with an action ${\cal S}$ depending on counting fields $\chi,\xi$, conjugated to the transferred charge and the energy of the island, respectively \cite{heikkila09,kindermann04}. For single-electron tunneling, the dynamics of a SET is governed by a master equation: The stationary probability distribution of charge states labeled by $N$, $p_N$, satisfies
\begin{equation}\label{eq:master}
 \sum_{N'}\Sigma_{N,N'}p_{N'}=0,
\end{equation}
where  elements of $\Sigma_{N,N'}$ correspond to single-electron tunneling rates so that $N'=N\pm 1$.  It is shown in the theory of full counting statistics (FCS)  \cite{bagrets03,kindermann04} that in order to obtain the action, one should modify ${\Sigma}_{N,N'}$ to include counting fields $\chi,\,\xi$. The action is then given by the eigenvalue of so-modified matrix with the smallest real part, $\sigma(\chi,\xi)$ \cite{bagrets03}. One could include higher-order tunneling processes by replacing  $\Sigma_{N,N'}$ with self-energies composed of all possible irreducible tunneling diagrams that take the SET from charge state $N'$ to $N$ \cite{schoeller94}. This is the way to account for cotunneling. Usually, if the tunneling processes of different orders become equally important, the situation is very difficult to comprehend \cite{nazarov09}. This situation typically occurs if the rate of electron transfer is comparable with the energy released in the course of transfer. This implies that the flow of charges can not be divided into separate events of any order. 

Fortunately, this is not the case of the fully overheated SET where cotunneling and SE events are separated even if they are equally important for transport. To understand this, let us concentrate on a blockaded diamond corresponding to a certain charge state, say $N=0$, and assume $T \ll eV_\mr{th}/k_\mr{B}$. The first SE transfer must be thermally activated and proceeds with the suppressed rate $\Gamma\exp(-W/k_\mr{B}T)$, $W=e(V_\mr{th}-V_\mr{b})/2$. It brings the island to the closest excited state $N=1$. However, the island will quickly get back to $N=0$: The first SE transfer is followed by a second, after a time given by an unsuppressed rate $\Gamma$. Similarly, a cotunneling event can also be viewed as two SE events separated by a time interval $\hbar/eV_\mr{th}$ \cite{nazarov09}. We see that the transport separates to elementary events each encompassing two SE transfers. The events are independent since time interval between them exceeds the time separation between the transfers by a large factor, $\min(G_\mr{Q}/G, \exp(W/k_\mr{B}T))$. Therefore, the cotunneling and SE contributions can simply be summed together, $\mathcal{S}=\mathcal{S}_\mr{se}+\mathcal{S}_\mr{cot}$.

Analytical results can be obtained by taking into account only two charge states on the island, $N=0$ and $N=1$. However, the validity of this approach requires $\log(G_\mr{Q}/G) \gg 1$, rarely the case in practical devices. Therefore we also perform accurate numerics, where we take more charge states for $\mathcal{S}_\mr{se}$ and weight $\mathcal{S}_\mr{cot}$ with the probabilities of those states.
\begin{figure}
 \includegraphics[width=\columnwidth]{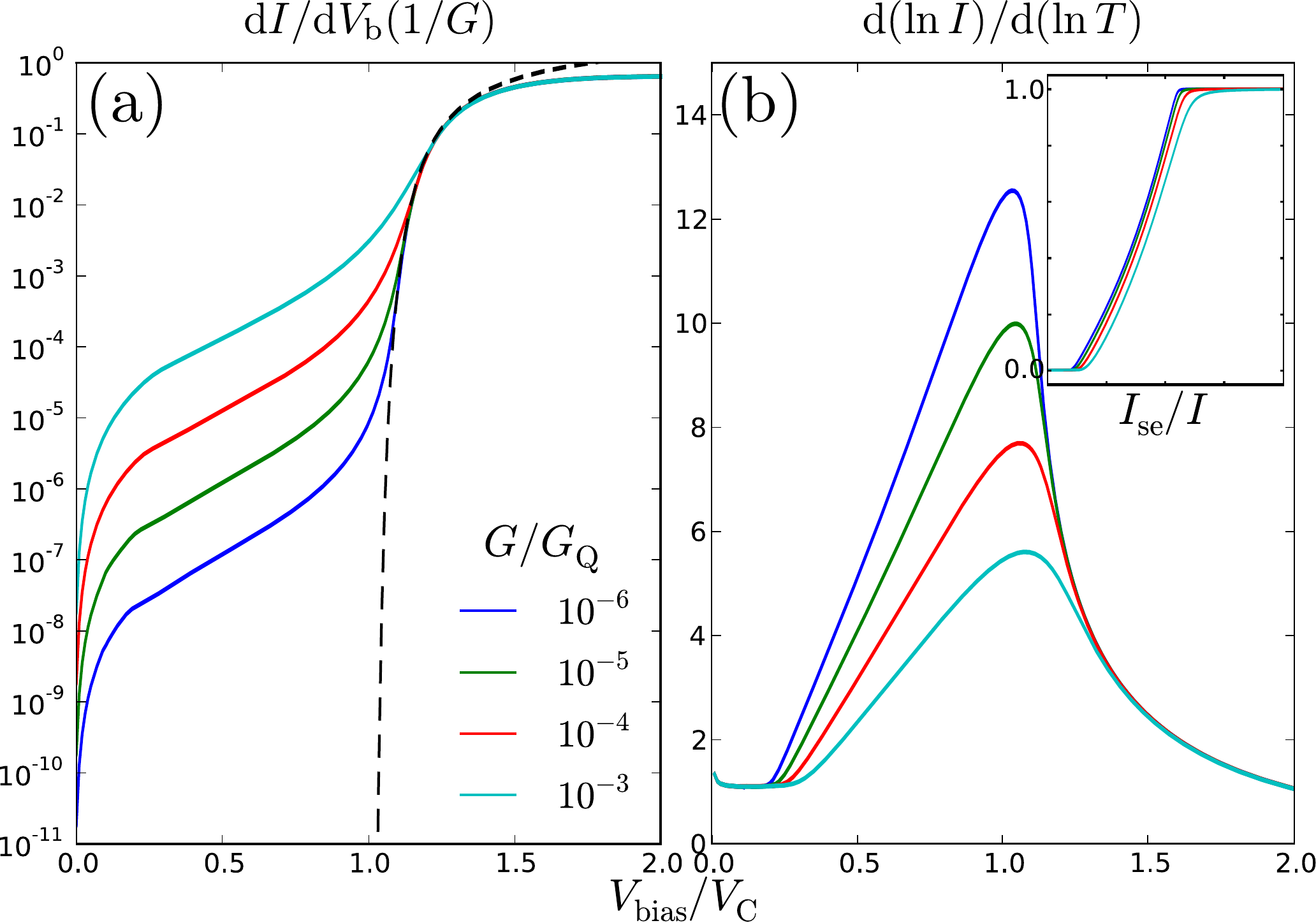}
 \caption{(color online) (a) Differential conductance as a function of $V_\mr{b}$ for several values of $G$. Dashed line is an asymptote for pure SE tunneling. (b) Temperature sensitivity $\d(\ln I)/\d(\ln T)$ as a function of $V_\mr{b}$ for the same values. Inset: Fraction of SE transfers in the current flow.}
 \label{fig:didv_teq}
\end{figure}

%regimes: SE
Let us first outline the three different regimes mentioned. To simplify the formulas, we set $e=k_\mr{B}=\hbar=1$ and define $\gamma=G/\pi G_Q$. The SE part of the action in the relevant limit $T \ll V_\mr{b},V_\mr{C};\;T\xi \ll 1$ reads (see Appendix)
\begin{align}
\label{eq:seqaction}
 {\cal S}_{\mr{se}}=&\gamma Te^{-W(T^{-1}+\xi)}\left\{\left[e^{-\chi}(1-e^{(W+V_\mr{b})\xi})\right.\right. \nonumber \\ &\left.\Bigl. +(1-e^{W\xi})\Bigr](\xi^{-1}-T)V_\mr{th}^{-1}+e^{W\xi}\right\}.
\end{align}
Using this action one evaluates the charge current, $I=\partial_\chi \mathcal{S}|_{\chi=0}$, and heat current, $\dot{H}=\partial_\xi \mathcal{S}|_{\xi=0}$, as functions of temperature. Equating the latter to zero yields the average temperature in the {\it SE regime}, $T=\frac{1}{4V_\mr{th}}\left(V_\mr{b}-V_\mr{C}\right)\left(V_\mr{b}+V_\mr{C}+2V_\mr{th}\right)$ above $V_\mr{C}$. The current steeply rises at the threshold, $I_\mr{se} \propto \exp(-W_\mr{C}/T) = \exp(-\frac{\sqrt{2}W_\mr{C}}{V_\mr{b}-V_\mr{C}})$ at $V_\mr{b}\approx V_\mr{C}$. 

%Cotunneling
The {\it cotunneling regime} takes place at $V_\mr{b} \ll V_\mr{C}$ where $T \approx V_\mr{b}$. In this region, the action reads (see Appendix)
\begin{align}\label{eq:cotaction}
 \mathcal{S}_\mr{cot}=&\gamma^2\left(\frac{1}{\delta E^+}+\frac{1}{\delta E^-}\right)^2\pi T^3\left[e^{-\chi}\mathcal{I}(v,x)+e^{\chi}\mathcal{I}(-v,x)\right. \nonumber \\ &\left.+2\mathcal{I}(0,x)-\mathcal{I}(v,0)-\mathcal{I}(-v,0)-2\mathcal{I}(0,0)\right],
\end{align}
where $\mathcal{I}(v,x)=\int_{-\infty}^{\infty}\d z\frac{e^{\ii v(z-\ii0^+)}}{4(z-\ii0^+)^2\sinh^2\left(z+\ii x\right)}$, $v=V_\mr{b}/\pi T$, and $x=\pi T\xi$. This yields the average temperature  $T\approx0.238V_\mr{b}$ and electric current $$I\approx 0.056 \gamma^{2}\left(\frac{1}{\delta E^+}+\frac{1}{\delta E^-}\right)^{2}V_\mr{b}^3.$$ Let us note that the current in the absence of overheating is also $\propto V_\mr{b}^3$ and is given by the same expression with coefficient $1/12\pi$. Therefore, the overheating enhances cotunneling current roughly by a factor of 2.

%Experimental
Thereby we resolve a long-standing discrepancy between theory and experiment. The pioneering work \cite{geerligs90} on cotunneling in SET has reported such a factor of 2 mismatch for the most conductive junctions. This is explained by full overheating. For the less conductive junctions the mismatch factors were $1.4$ and $1.2$, explained by incomplete overheating. In this case the smaller electronic heat flows may have become comparable with phonon heat transfer. In addition, \refcite{geerligs90} reports a crossover to SE regime at approximately half of the $V_\mr{th}$ expected: this conforms to theoretical value of $V_\mr{C}$.

%Competition regime
A further increase of $V_\mr{b}$ increases the temperature and activates SE processes. Comparing cotunneling ($\sim \gamma V^3_\mr{b}/W^2$) and SE ($\sim \gamma T \exp(-W/T)$) rates, we expect the SE processes to become important at $V_\mr{b} \sim T \sim W \ln(1/\gamma) \ll V_\mr{C}$. We enter the regime where equilibrium temperature is determined from the competition between SE transfers cooling the island and cotunneling events heating the island. Similar estimation gives with logarithmic accuracy $T\approx W/\ln(1/\gamma)$ in the whole interval of the competition, that is, up to $V_\mr{C}$. Inset in Fig.~\ref{fig:didv_teq}(b) shows the fraction of SE events in the current flow. The fraction grows almost linearly from the border of the cotunneling regime up to $V_\mr{C}$. Indeed, at low voltages each cotunneling process provides $V_\mr{b}$ of heat while a SE transfer cools the island by a value of $\sim W$: Many cotunneling events match up a single SE transfer. Near $V_\mr{C}$, a SE transfer gives a vanishing cooling $\sim(V_\mr{C}-V_\mr{b})$: Many SE transfers are needed to balance a single cotunneling.  A simple analytical expression for the total current in this regime does not exist. Qualitatively, it is estimated by $I = (1+ (1-V_\mr{b}/V_\mr{C})^{-1}) I_\mr{cot}$, $I_\mr{cot} \propto \gamma^2$ being the cotunneling current in the absence of overheating.

%Temperature sensitivity 

It is interesting to note an anomalously high temperature sensitivity of the fully overheated SET. The underlying mechanism is the same as in living organisms that rely on the balance of the thermally activated rates $\sim\exp(-W/T)$. Changing a rate by a factor of $e$ is achieved by a small temperature change $\delta T \simeq T (T/W) \ll T$ and may even lead to the destruction of an organism. Similarly, the thermally-activated character of SE transfers may result in a high sensitivity of the current that we characterize by a dimensionless number $\d(\ln I)/\d(\ln T) \simeq W/T$. This is plotted in Fig.~\ref{fig:didv_teq}(b). We see that the sensitivity reaches the maximum at the crossover between competition regime where it indeed scales as $W_\mr{C}/T_\mr{C}$. The linear growth below $V_\mr{C}$ is explained by the almost linear increase of the fraction of SE electron transfers in the competition regime. Above $V_\mr{C}$, the sensitivity drops like $1/(V_\mr{b}-V_\mr{C})$ owing to temperature increase.

\begin{figure}
 \includegraphics[width=\columnwidth]{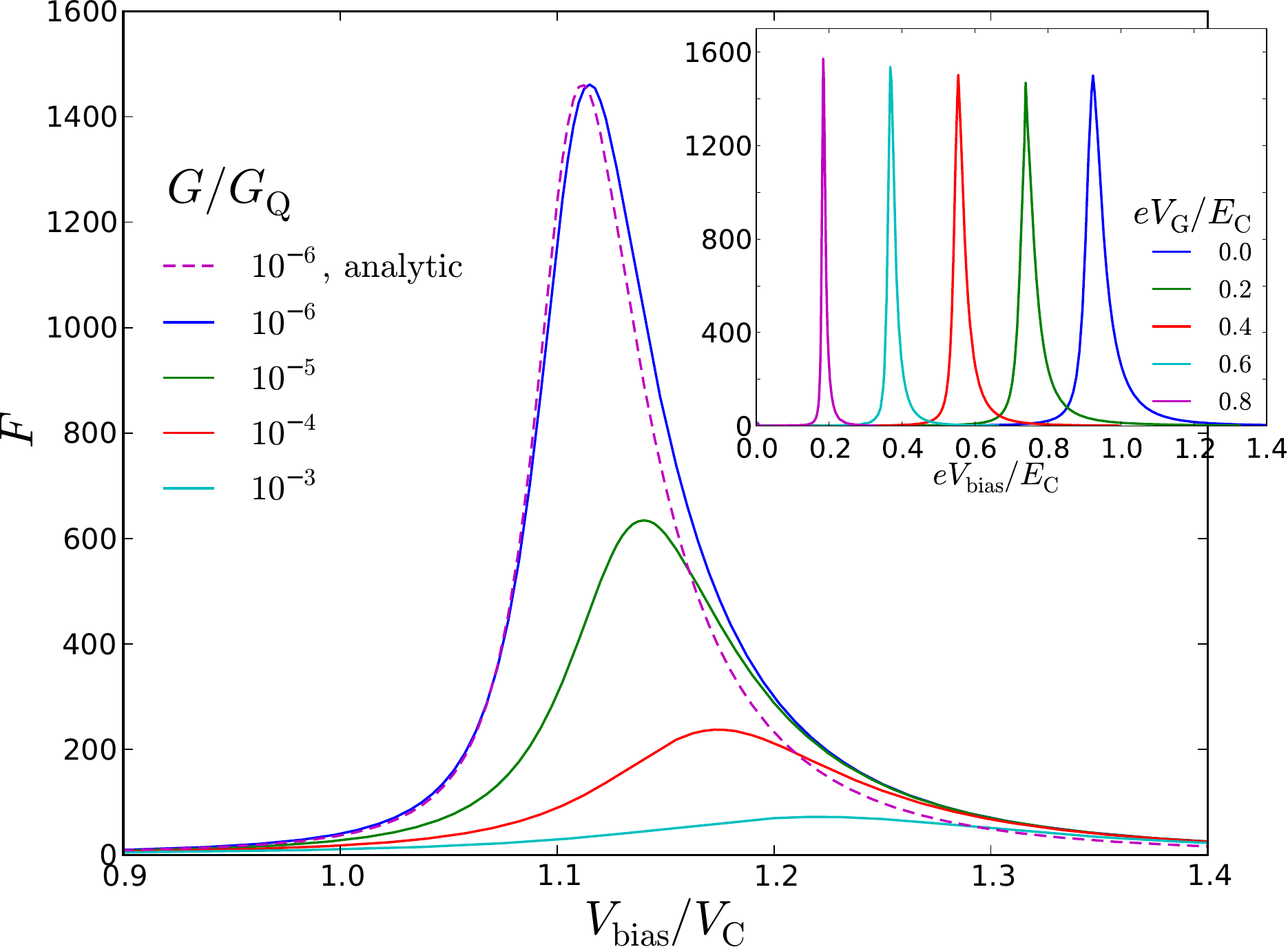}
 \caption{(color online) Fano factor of the temperature fluctuation induced current noise for different values of $G/G_\mr{Q}$. Dashed line is an analytic approximation for $G/G_\mr{Q}=10^{-6}$, and agrees well with the numerical result. Inset shows the Fano factor for $G/G_\mr{Q}=10^{-6}$ in an absolute voltage scale for various values of $eV_\mr{G}/E_\mr{C}$. The peaks fall on top of each other once rescaled to common $V_\mr{C}$}
 \label{fig:f_tfluct}
\end{figure}

%Temperature fluctuations and Fano factor
Discreteness of charge transfer through the structure gives rise to a current noise, $S_I$, and heat current noise, $S_{\dot{H}}$, both white at frequencies $\omega\ll I$. The heat current noise produces temperature fluctuations that persist over a significant time, $\tau=(\partial \mathcal{F}/\partial T)/(\partial \dot{H}/\partial T)$ ($\mathcal{F}$ being the total free energy of the island, proportional to its volume). The fluctuation of temperature is given by $\langle (\Delta T)^2\rangle = S_{\dot{H}}/\tau(\partial \dot{H}/\partial T)^2$. Temperature fluctuations change the current, $\delta I = (\partial I/\partial T) \delta T(t)$ giving rise to extra ``slow'' current noise persisting at frequencies $\simeq 1/\tau$,
\begin{equation}
 S_{I,\mr{slow}}=\left(\frac{\partial I/\partial T}{\partial \dot{H}/\partial T}\right)^2 S_{\dot{H}},
\end{equation}
manifesting the overheating. Anomalous temperature sensitivity gives rise to an anomalous Fano factor, plotted in Fig.~\ref{fig:f_tfluct}. Similar to sensitivity, the Fano factor also peaks at the crossover between SE and competition regimes.

%rescaling
Let us now concentrate on the crossover region. Three factors contribute to the heat balance at $V_\mr{b}\approx V_\mr{C}$: 
\begin{equation}
0 = \dot{H}_\mr{cot} + 2^{-1/2}\gamma  T e^{-W/T}(V_\mr{b}-V_\mr{C})  - \gamma T^2 e^{-W/T},
\end{equation}
cotunneling heating $\dot{H}_\mr{cot}\propto\gamma^2V_\mr{C}^2$ that is approximately constant, SE flow that switches from cooling to heating at $V_\mr{b}=V_\mr{C}$, and extra cooling that stabilizes temperature in SE regime. We define the temperature $T_\mr{C}$ at the crossover through
$$
 \frac{\dot{H}_\mr{cot}V_\mr{C}}{\sqrt{2}\gamma T_\mr{C}^3} 
 \exp\left(\frac{ V_\mr{C}}{\sqrt{2}T_\mr{C}}
 -\frac{1}{\sqrt{2}}\right)=1\Rightarrow \frac{T_\mr{C}}{V_\mr{C}} \approx 
 \frac{2^{-1/2}}{\ln(1/\gamma)},  
$$
and introduce dimensionless deviations of voltage $x$ and temperature $y$ such that $V_\mr{b}=V_\mr{C}+\sqrt{2}T_\mr{C}+2xV_\mr{C}(T_\mr{C}/V_\mr{C})^2$ , $T=T_\mr{C}+\sqrt{2}yV_\mr{C}(T_\mr{C}/V_\mr{C})^2$, $(T_\mr{C}/V_\mr{C})$ being an important dimensionless small parameter enabling the scaling. The heat balance rescales to 
\begin{equation}
 e^y(x-y) + 1=0,
\end{equation}
which implicitly gives the temperature as a function of voltage. The crossover takes place at $x,y \approx 1$, the rescaled equation being valid in a larger interval up to $x,y \approx V_\mr{C}/T_\mr{C}$. We see that the crossover is shifted from $V_\mr{C}$ by $\sqrt{2}T_\mr{C}$. The width of crossover interval in voltage/temperature is small $\sim  T_\mr{C}(V_\mr{C}/T_\mr{C}) \ll T_\mr{C}$. While temperature changes insignificantly, the relative change of quantities of interest is by an order of magnitude. Current is rescaled to
\begin{equation}
 I(y)=\frac{\gamma T}{\sqrt{2}}\exp\left(-\frac{V_\mr{th}-V_\mr{b}}{2T}\right)=\frac{\dot{H}_\mr{cot}}{2 V_\mr{C}}\left(\frac{V_\mr{C}}{T_\mr{C}}\right)^2e^y.
\end{equation}
Similarly, 
\begin{align}
 S_{\dot{H}}&=\frac{1}{3}V_\mr{C}^2I(y),\;\frac{\partial I}{\partial T}=\frac{1}{\sqrt{2}V_\mr{C}}\left(\frac{V_\mr{C}}{T_\mr{C}}\right)^2I(y), \nonumber \\ 
 \frac{\partial\dot{H}}{\partial T}&=\sqrt{2}(1+e^{-y}) I(y). 
\end{align}
%Finally Fano
This yields the Fano factor at the crossover
\begin{equation}
 F=\frac{(\partial I/\partial T)^2}{(\partial\dot{H}/\partial T)^2}\frac{S_{\dot{H}}}{I}=\frac{1}{12}\left(\frac{V_\mr{C}}{T_\mr{C}}\right)^{4}\frac{1}{\left(1+e^{-y}\right)^{2}}, 
\end{equation}
describing a sharp rise as $y\to0$. Its fall in the SE region is described by substituting $T_\mr{C} \to T(V_\mr{b})$. This fits well to the numerical results as shown in Fig.~\ref{fig:f_tfluct}.

There are very interesting statistics of temperature fluctuations at the crossover. Generally \cite{heikkila09}, one expects deviations from Gaussian statistics for temperature deviations of the order of average temperature that occur with exponentially small probability $\ln P \simeq -\bar{T}/\delta_S$, $\delta_S$ being the single-electron level spacing in the island. In an overheated SET around the crossover, the deviations $\delta T \simeq T_\mr{C}(T_\mr{C}/V_\mr{C})$ are already non-Gaussian and their probability is greatly enhanced, $\ln P \simeq - (T_\mr{C}/\delta_S)(T_\mr{C}/V_\mr{C})^4$. We will present the detailed results for the distribution of the fluctuations in a separate publication. The system ideally suits for the experimental observation of temperature fluctuations at nanoscale: large fluctuations are easily read by a measurement of the electric current, slowly fluctuating in time.

Finally, let us estimate the importance of electron--phonon interaction to assess the feasibility of full overheating. For such an estimate, it is enough to add a term to the action $\mathcal{S}_{e-ph}=-\Sigma\mathcal{V}T^5\xi(1-5 T\xi)$, where $\mathcal{V}$ is the volume of the island and $\Sigma$ the material-specific electron--phonon coupling constant.
The island is overheated provided $\Sigma\mathcal{V}T^3\ll\gamma$. For typical values \cite{giazotto06}, $\Sigma\approx2\times10^{-9}\:\mr{W}\mr{K}^{-5}\mu\mr{m}^{-3}$, $T\approx0.1\:\mr{K}$, and $\gamma\approx10^{-3}$, the volume of the island should be of the order of $\mathcal{V}\approx10^{-4}\:\mu\mr{m}^3$ to reach this regime. This is easily achievable experimentally. High Fano factor $\approx 10^3$ requires $\gamma\approx10^{-5}$ and $\mathcal{V}\approx10^{-6}\:\mu\mr{m}^3$, feasible in smaller systems such as granular metals and multi-walled carbon nanotubes.

To conclude, we have studied the fully overheated SET revealing the importance of cotunneling processes that compete with single-electron transfers in a wide interval of bias voltages. The fully overheated SET exhibits anomalous temperature sensitivity and ``slow'' current noise with a huge Fano factor as a result of temperature fluctuations. These effects are most pronounced at the crossover between competition and single-electron tunneling dominated regimes.

M.A.L.~acknowledges the support from the Finnish Academy of Science and Letters, and T.T.H.~the support from the Academy of Finland.

\bibliography{fcs}

\begin{widetext}
\appendix

\section{Derivation of Eq.~(2)}

Let us define a function
\begin{equation} \label{eq:gamma}
 \tilde{\Gamma}^\alpha(\chi,\Delta)\equiv\gamma^\alpha e^{-\chi}\frac{Te^{-\Delta(\xi+1/T)}}{1+T\xi}{}_2F_1(1,1+T\xi;2+T\xi;-e^{-\Delta/T}),
\end{equation}
where $_2F_1(a,b;c;z)$ is the Gauss hypergeometric function. The counting field modified single-electron tunneling rates at the left junction are
\begin{equation*}
 \tilde{\Gamma}^L_{N+1,N}=\tilde{\Gamma}^L(\chi_L-\chi_D,\delta E^+_N-\mu_L+\mu_D),\quad\tilde{\Gamma}^L_{N-1,N}=\tilde{\Gamma}^L(\chi_D-\chi_L,\delta E^-_N+\mu_L-\mu_D),
\end{equation*}
and similarly for the right junction. When $T\ll|\Delta|$ we have
\begin{align} \label{eq:approx2f1}
 {}_2F_1(1,1+T\xi;2+T\xi;-e^{-\Delta/T})&\approx\left\{\begin{array}{cc}
                                                        1-\frac{1+T\xi}{2+T\xi}e^{-\Delta/T}, & \Delta>0 \\
                                                        e^{\Delta/T}\left[e^{\Delta\xi}\Gamma(-T\xi)\Gamma(2+T\xi)+\frac{\Gamma(T\xi)\Gamma(2+T\xi)}{\Gamma(1+T\xi)^2}\right], & \Delta<0
                                                       \end{array}\right. \\ \label{eq:approxgamma}
 \Rightarrow\tilde{\Gamma}^\alpha(\chi,\Delta)&\approx\left\{\begin{array}{cc}
                                                        \gamma^\alpha e^{-\chi}\frac{Te^{-\Delta(\xi+1/T)}}{1+T\xi}, & \Delta>0 \\
                                                        \gamma^\alpha e^{-\chi}\frac{e^{-\Delta\xi}-1}{\xi}, & \Delta<0
                                                       \end{array}\right..
\end{align}
Here $\Gamma(z)$ is the gamma function.

At low temperatures it is enough to consider two charge states, $N=0$ and $N=1$, on the island. The effective action is then easy to calculate, yielding
\begin{equation}
 \mathcal{S}=\frac{1}{2}\left[\Gamma_{0,1}+\Gamma_{1,0}-\sqrt{(\Gamma_{0,1}+\Gamma_{1,0})^2+4(\tilde{\Gamma}_{0,1}\tilde{\Gamma}_{1,0}-\Gamma_{0,1}\Gamma_{1,0})}\right]\approx-\frac{\tilde{\Gamma}_{0,1}\tilde{\Gamma}_{1,0}-\Gamma_{0,1}\Gamma_{1,0}}{\Gamma_{0,1}+\Gamma_{1,0}}.
\end{equation}
The last approximation is valid when $\Gamma_{0,1}\gg\Gamma_{1,0}$ or $\Gamma_{0,1}\ll\Gamma_{1,0}$, i.e., under Coulomb blockade. Using the form of Eq.~\eqref{eq:approxgamma} this becomes (for $\chi_L\equiv\chi$, $\chi_D=\chi_R=0$, $\gamma^L=\gamma^R=\gamma$, $\mu_L=-\mu_R=eV_\mr{b}/2$, and $\delta E^+=E_\mr{C}-eV_\mr{G}$)
\begin{align}
 \mathcal{S}=&-\frac{\gamma Te^{-\delta E^+/T}}{\xi(1+T\xi)\delta E^+}\frac{1}{1+\frac{T}{\delta E^+}e^{-\delta E^+/T}\cosh\left(\frac{eV_\mr{b}}{2T}\right)}\times \nonumber \\ &\left\{\cosh\left[\frac{eV_\mr{b}}{2}\left(\frac{1}{T}+2\xi\right)-\chi\right]-e^{-\delta E^+\xi}\cosh\left[\frac{eV_\mr{b}}{2}\left(\frac{1}{T}+\xi\right)-\chi\right]\right. \nonumber \\ &\left.-e^{-\delta E^+\xi}\cosh\left[\frac{eV_\mr{b}}{2}\left(\frac{1}{T}+\xi\right)\right]+\cosh\left(\frac{eV_\mr{b}}{2T}\right)\left[1-2\delta E^+\xi(1+T\xi)\right]\right\}.
\end{align}
This results in Eq.~(2) of the main text in the limit $T \ll V_\mr{b},V_\mr{C};\;T\xi \ll 1$.

\section{Derivation of Eq.~(3)}

Cotunneling rate with the counting fields between leads $\alpha$ and $\beta$ ($\in\{L,R\}$) held at zero temperature
\begin{align}
 \tilde{\Gamma}^{\beta\alpha}_\mr{cot,N}=&\frac{1}{2\pi}\gamma^\alpha\gamma^\beta e^{-\chi}\int_{-\infty}^{\infty}\d E\left(\frac{1}{E+eV_\mr{b}/2-\delta E^+_N}+\frac{1}{-E-eV_\mr{b}/2-\delta E^-_N}\right)^2\times \nonumber \\
 &\int_{-\infty}^0\d\epsilon[1-f(\epsilon-E)]e^{\xi(\epsilon-E)}\int^{\infty}_0\d\epsilon'f(\epsilon'-E-eV_\mr{b})e^{-\xi(\epsilon'-E-eV_\mr{b})},
\end{align}
where $f(E)$ is the Fermi function at the island temperature and $eV_\mr{b}=\mu_\alpha-\mu_\beta$. Now write
\begin{equation}
 [1-f(\epsilon-E)]e^{\xi(\epsilon-E)}=\int_{-\infty}^{\infty}\d t F_1(t)e^{\ii(\epsilon-E)t},\quad f(\epsilon'-E-eV_\mr{b})e^{-\xi(\epsilon'-E-eV_\mr{b})}=\int_{-\infty}^{\infty}\d t' F_2(t')e^{\ii(\epsilon'-E-eV_\mr{b})t'},
\end{equation}
and assume $eV_\mr{b}\ll\delta E^+_N,\delta E^-_N$ so that
\begin{equation}
 \tilde{\Gamma}^{\beta\alpha}_\mr{cot,N}=-\gamma^\alpha\gamma^\beta e^{-\chi}\left(\frac{1}{\delta E^+_N}+\frac{1}{\delta E^-_N}\right)^2\int_{-\infty}^{\infty}\d t\frac{F_1(t)F_2(-t)e^{\ii eV_\mr{b}(t-\ii0)}}{(t-\ii0)^2}.
\end{equation}
Carrying out the Fourier transforms gives us
\begin{equation}
 F_1(t)=\frac{-\ii T}{2\sinh(\pi T(t+\ii\xi))},\quad F_2(t)=\frac{\ii T}{2\sinh(\pi T(t-\ii\xi))},
\end{equation}
resulting in
\begin{equation}
 \tilde{\Gamma}^{\beta\alpha}_\mr{cot,N}=\gamma^\alpha\gamma^\beta e^{-\chi}\left(\frac{1}{\delta E^+_N}+\frac{1}{\delta E^-_N}\right)^2\pi T^3\int_{-\infty}^{\infty}\d z\frac{e^{\ii v(z-\ii0^+)}}{4(z-\ii0^+)^2\sinh^2\left(z+\ii x\right)},
\end{equation}
where $v=eV_\mr{b}/\pi T$ and $x=\pi T\xi$. This leads to Eq.~(3) of the main text when the four different processes, corresponding to cotunneling from the left lead to the right lead, right to left, left to left, and right to right, are summed and the $\chi=\xi=0$ part is subtracted. The integral above can be evaluated by closing the contour around the upper (lower) half plane for $v>0$ ($v<0$). The result is
\begin{align}
 \mathcal{I}(v>0,x)&=\frac{\pi}{2}\frac{v+2\cot x}{\sin^2 x}+\frac{\pi}{2}\sum_{n=0}^{\infty}\frac{e^{vx-n\pi v}(vx-n\pi v-2)}{(x-n\pi)^3}, \\
 \mathcal{I}(v<0,x)&=-\frac{\pi}{2}\sum_{n=1}^{\infty}\frac{e^{vx+n\pi v}(vx+n\pi v-2)}{(x+n\pi)^3}.
\end{align}
These sums can be used to evaluate the action to a desired accuracy.
\end{widetext}

\end{document}